\documentclass[twocolumn,showpacs,superscriptaddress,pra]{revtex4-1}
\usepackage{graphicx}
\usepackage{times}
\usepackage{textcomp}
\usepackage{amsmath,amssymb}
\usepackage{color}
\newcommand{\+}[1]{\ensuremath{\boldsymbol{#1}}}
\begin{document}
\title{Three-level spin system under  decoherence-minimizing  driving fields:
Application to  nitrogen-vacancy spin dynamics}
\author{S. K. Mishra}
\affiliation{ Department of Physics, Indian Institute of Technology, Banaras Hindu University, Varanasi - 221005, India}
\author{L. Chotorlishvili}
\affiliation{Institut f\"{u}r Physik, Martin-Luther-Universit\"{a}t Halle-Wittenberg, 06120 Halle, Germany}
\author{A. R. P. Rau}
\affiliation{Department of Physics and Astronomy, Louisiana State University, Baton Rouge,
Louisiana 70803-4001, USA}
\author{J. Berakdar}
\affiliation{Institut f\"{u}r Physik, Martin-Luther-Universit\"{a}t Halle-Wittenberg, 06120 Halle, Germany}
\begin{abstract}
Within the framework of a general three-level problem, the dynamics of the nitrogen-vacancy (NV) spin is studied
for the case of a special type of external driving consisting of  a set of continuous  fields with decreasing intensities. Such a set has been proposed for minimizing coherence losses. Each new driving field with smaller intensity is designed to protect against the fluctuations induced by the driving field at the
preceding step with larger intensity. We show that indeed this particular type of external driving minimizes the loss of coherence, using purity and entropy as quantifiers for this purpose. As an illustration, we study the coherence loss of an NV spin due to a surrounding spin bath of $^{13}$C nuclei.
 \end{abstract}
\date{\today }

\maketitle
 Three-level systems arise in many physical contexts. A particularly interesting case
 is the effective Hamiltonian of the electronic spin with three states, $|m_S\rangle =|-1\rangle ,~|0\rangle ,~|1\rangle $, of a nitrogen vacancy (NV) center in diamond that serves currently as a platform
for addressing a variety of fundamental questions and applications in quantum information  (see, for example, Refs.~\cite{8,9,10,11,12,13,14,15,16,17,rabl,zhou,18,19,20,21,22,23,24,25,26,27,28,29,30}).
Field-free NV centers in diamond
possess very long decoherence times even at room temperature.  On the other hand, a precise control of the dynamics of NV spins  is achievable via external driving fields
rendering possible the use of NV impurity spins as sensors in magnetic resonance force microscopy \cite{rugar,su}.

For practical applications, particularly in quantum information processing, the loss
of coherence is a critical issue and ways to  minimize it are of vital importance. In the last few years, the protection of the state of the system from the environment has been addressed by
 applying fast and strong pulses \cite{vola,du,delange,ryan,souza,biercuk}. The motivation behind such pulsed dynamical decoupling is attributed
to Hahn's 'spin-echo' experiment where refocusing pulses are used to decouple the spin and the environment \cite{hahn}. However, spin-echo is related to the collective effect for a macroscopic number of spins and thus not directly applicable to quantum logic gates.

The basic problem for quantum logic gates is commutation between the decoupling pulses and the control pulses. In spite of the success of pulsed dynamical decoupling schemes in decoupling the deleterious effects of environment, their role has been found to be complicated when implemented with quantum logic gates \cite{vadersar}.  NV centers usually are surrounded by $^{13}$C nuclear spins and cannot be completely decoupled from the unwanted interactions with that $^{13}$C spin-based environment. The resulting loss of  coherence of the central NV spin has been a serious concern for the usage of NV centers in quantum computing. The environment acts effectively as a random magnetic field which contributes to the energy level splitting of the NV center, leading to decoherence. Therefore, a challenge for any protection scheme for NV centers is to minimize the effect of this random magnetic field.

In order to achieve flexibility in implementing various high fidelity quantum logic tasks, a continuous wave dynamical decoupling 
scheme has been introduced recently \cite{rabl,fanchini,lidar,bermudez,bermudez1,timoney,xu} with the following
basic idea: An equally weighted superposition of the states $|-1\rangle ,~|1\rangle $ can result in states with eigenvalues insensitive to the random field. Such an equally weighted superposition can be obtained by applying two off-resonant continuous microwave driving fields to $|0\rangle \rightarrow|1\rangle ,~|0\rangle \rightarrow|-1\rangle $. 
This scheme is easily realizable  experimentally and was successful in controlling  to some extent the
decoherence  of quantum states due to the environment. While protecting the NV center from the influence of the environment,
  this scheme  has however
  collateral effects as well, itself being a source of another type of decoherence identified with unavoidable fluctuations of the amplitude of the driving pulses. Random and systematic fluctuations arising from the driving field (microwave) may reduce the coherence time of the quantum state of the NV spins.

In a recent paper, Cai {\it et al.} \cite{cai} have considered a particular type of external driving scheme termed as concatenated continuous decoupling scheme. This scheme is designed to minimize the loss of coherence
due to the environment and also from the fluctuations of the amplitude of driving field.
The main idea of
 their proposed driving protocol is to provide a set of
continuous driving fields with decreasing intensities. A smaller intensity of the driving field
naturally leads to a lower absolute value of fluctuations.
 In such a driving protocol, each new driving field with a smaller intensity is supposed
 to protect against the fluctuations caused by the driving field acting at the
 preceding step with a larger intensity. This driving protocol is quite complicated, however,
 for practical applications  in the case of a three-level system. Other papers \cite{li,bogoliubov} have, therefore, addressed this issue
for a three-level system only for a simpler driving case.

To go further, we consider a method that has been developed to study the time dependence of a general three-level system \cite{rau1,rau2,rau3} which, in principle, allows the study of arbitrary
driving. In the present work, we utilize this method \cite{rau1,rau2,rau3} to study the NV three-level spin system and consider the
non-monochromatic external driving scheme proposed
by Cai {\it et al.} \cite{cai}. Our ultimate goal is to minimize the loss of coherence between the NV spin system and the driving field.
We investigate the loss of coherence due to neighbouring $^{13}$C nuclear spins and check the applicability of non-monochromatic external driving schemes to minimize the loss. For this purpose, we quantify the coherence between the spin system and the driving field in terms of purity \cite{schleich} and entropy.
\section{Model}
\label{model}
The NV center in diamond consists of a substitutional nitrogen atom with an
adjacent vacancy. The total spin of the many-electron orbital ground state of the NV center is
described by the spin triplet $S = 1, m_S= -1, 0, 1$. States with different $m_S$ are separated by a
 zero-field splitting which is of the order $\omega_0 = 2.88 \rm{GHz}$ \cite{rabl}. This kind of splitting is intrinsic to the
NV spin system, originating from spin-spin interactions leading to the single-axis
spin anisotropy $DS_z^2 \sim \omega_0$ \cite{wrachtrup}, where $S_z$ is the component of spin along the quantization axis. Throughout this paper, we set $\hbar=1$.
When an external magnetic
field $\vec{B}_0$ is applied such that $\mu_B\vert\vec{B}_0\vert < \omega_0$, the degeneracy is removed due to the Zeeman shift of the levels $\vert-1\rangle$ and $\vert1\rangle$. The amplitude of the Zeeman shift is proportional to the term $\sim B_0 S_z$.
Microwave fields drive Rabi oscillations between ground $|0\rangle $ and the excited states $|\pm\rangle $.

\begin{figure}[!t]
\includegraphics [angle=0,width=.9\columnwidth] {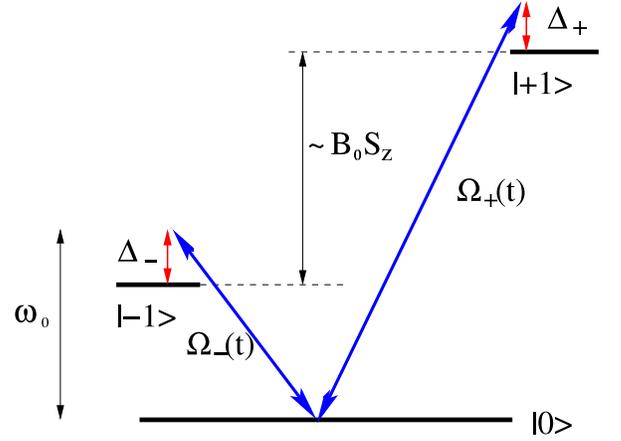}
\caption{(color online) A level diagram for the NV center with time-dependent driving fields. }
\label {three_lev}
\end{figure}

The Hamiltonian of
a single NV spin system is given by:
\begin{equation}
 H_{NV}=\sum_{i=\pm1}\biggl(-\Delta_i\vert i\rangle\langle i \vert+\frac{\Omega_i}{2}\bigl(\vert 0 \rangle\langle i
\vert + (\vert i \rangle\langle 0 \vert \bigr)\biggr).
\label{ham}
\end{equation}
Here $\Delta_{i}$ and $\Omega_{i}$ denote the detunings and Rabi frequencies of the two microwave transitions.
For a weak magnetic field such that $\mu_B\vert\vec{B}_0\vert \ll \omega_0$, one can neglect the level splitting and set
$\Delta_{-1}=\Delta_{+1}$, and similarly for the driving fields, $\Omega_{-1} = \Omega_{+1}$. In this case, the Hamiltonian in Eq.~(\ref{ham}) couples the ground state $\vert 0\rangle$ to the ``bright''
superposition of the excited states, $\vert b\rangle=\frac{1}{\sqrt{2}}(\vert-1\rangle+\vert1\rangle)$. Therefore, the model is equivalent to an
effective two-level system. However, if the external magnetic field is strong enough, then we have to
proceed with a three-level model in Eq.~(\ref{ham}). A level diagram for the NV spin system is shown in Fig.~\ref{three_lev}.
In the basis $\vert0\rangle$, $\vert-1\rangle$ and $\vert 1\rangle$, the Hamiltonian in Eq. (\ref{ham}) can be rewritten as
(to simplify notation we abbreviate  $\Omega_{\pm 1}\equiv \Omega_{\pm};\, \Delta_{\pm 1}\equiv \Delta_{\pm}$)
\begin{equation}
 H=\begin{pmatrix}
  0 & \frac{\Omega_{-}}{2} & \frac{\Omega_{+}}{2} \\
  \frac{\Omega_{-}}{2}& -\Delta_{+} & 0 \\
  \frac{\Omega_{+}}{2} & 0 &-\Delta_{+}
 \end{pmatrix}.
\end{equation}

For fixed values of $\Omega_{-}$ and $\Omega_{+}$, the above model Hamiltonian is time independent and its time evolution can be easily calculated.
However, the solution is nontrivial for a time-dependent Hamiltonian with parameters $\Omega$ and $\Delta$ time varying functions. In what follows, we will consider a special time-dependent concatenated continuous decoupling (CCD) driving scheme which not only suppresses the decoherence due to the environment but also nullifies the effects of fluctuations in the driving field itself. This special driving has been implemented in experiments recently and the performance in improving the coherence time of the state of the
system against environmental fluctuations has been found to be exceptional.

Introduced by Cai {\it et al.} \cite{cai}, this special CCD driving scheme can control the coherence time of the NV spin by protecting it
against fluctuations of the driving field at the preceding level. Before using
this CCD scheme for the three-level NV centers, we will briefly discuss its implementation using a simpler two-level system.
Consider a two-level system given by the following Hamiltonian
\begin{eqnarray}
 \mathcal{H}=\mathcal{H}_0+\mathcal{H}_d^1,
\end{eqnarray}
where the first term $\mathcal{H}_0=\frac{\omega}{2}\sigma_z$ is the Hamiltonian in the absence of the driving field. It describes a system with two eigenstates $\vert\uparrow\rangle$ and $\vert\downarrow\rangle$ (eigenstates of Pauli matrix $\sigma_z$)
with eigenvalues $\omega_{\uparrow}(=+\omega/2)$ and $\omega_{\downarrow}(-\omega/2)$.
But environmental noise may cause fluctuations in $\omega_{\uparrow}$ and $\omega_{\downarrow}$. In an attempt to protect the state of the system against such fluctuations, a driving field term in the Hamiltonian $\mathcal{H}_d^{1}=\Omega(t)\sigma_x$ on resonance with the energy difference $\omega$ between the two eigenstates is introduced. In the
simplest case, we choose a continuous wave periodic driving field given by
\begin{eqnarray}
 \Omega(t)=\Omega_1\cos(\omega t).
\end{eqnarray}

In the interaction picture with respect to $\mathcal{H}_0$, we find $\mathcal{H}_I=e^{i\mathcal{H}_0 t}\mathcal{H}_d^1 e^{-i\mathcal{H}_0 t}$ and, subsequently
using the rotating wave approximation, $\mathcal{H}_I=\frac{\Omega_1}{2}\sigma_x$. The new eigenstates
$|\rightarrow\rangle_x=$ $\frac{1}{\sqrt{2}}(\vert\uparrow\rangle+\vert\downarrow\rangle)$ and $\vert\leftarrow\rangle_x=\frac{1}{\sqrt{2}}(\vert\uparrow\rangle-
\vert\downarrow\rangle)$ are dressed states and eigenstates of Pauli matrix $\sigma_x$. Thus the effect of fluctuation is only
to induce transitions among the dressed states. In this formalism,
the effects of fluctuations arising due to environment are suppressed. However, in realistic experiments, the fluctuations arising due to the
limited stability of the microwave source will still cause fluctuations in the energies of the dressed states. This fluctuation can be taken as a stochastic noise term in the amplitude of the first-order term.

In order to achieve protection from the
fluctuations arising due to the driving field, a concatenated set of continuous driving fields can be used to decouple the system with these fluctuations. The fluctuation
due to the driving field up to the first order term $\Omega_1\cos(\omega t)$ can be suppressed by introducing a second order term such that
\begin{eqnarray}
\Omega(t)= \Omega_1\cos(\omega t)+2\Omega_2\cos\biggl(\omega t+\frac{\pi}{2}\bigg)\cos\big(\Omega_{1} t\bigr).
\label{2nd_field}
\end{eqnarray}
This second-order term in the driving field is in resonance with the energy gap between dressed states due to the
first-order term. Thus, in the above equation Eq.~(\ref{2nd_field}), the first term protects the system from the fluctuations arising due to the surroundings while the second term minimizes the
fluctuations due to the first term. The amplitude of the second-order driving, $\Omega_2$ is smaller than $\Omega_1$.
In appendix \ref{append1}, we explicitly show how the effects of fluctuations caused by environment and driving fields simultaneously
are reduced by using the CCD driving.

We turn next to the more general case with level splitting not neglected and two driving fields distinguished by $\pm$ as in Fig.~\ref{three_lev}.
 In each driving field, we consider as above {\it first-order}, and {\it second-order}. These are,
\begin{eqnarray}
 \Omega_{\pm}^{(I)}(t)=\Omega_{\pm}^{(1)}\cos(\omega_{\pm} t),
\label{1st_order}
\end{eqnarray}
and
\begin{eqnarray}
\Omega_{\pm}^{(II)}(t)&=&\Omega_{\pm}^{(1)}\cos(\omega_{\pm} t)+2\Omega_{\pm}^{(2)}\cos\bigg(\omega_{\pm} t+\frac{\pi}{2}\bigg)\nonumber\\
&\times&\cos\bigg(\Omega_{\pm}^{(1)} t\bigg).
\label{2nd_order}
\end{eqnarray}
Similarly we may consider {\it third-order} driving fields as
\begin{eqnarray}
\Omega_{\pm1}^{(III)}(t)&=&\Omega_{\pm1}\cos(\omega_{\pm} t)+2\Omega_{\pm2}\cos(\omega_{\pm} t+\frac{\pi}{2})\cos(\Omega_{\pm1} t)\nonumber \\
&+&2\Omega_{\pm3}\cos(\omega_{\pm} t)\cos(\Omega_{\pm2} t).
\label{3rd_order}
\end{eqnarray}

In Eqs. (\ref{1st_order}), (\ref{2nd_order}) and (\ref{3rd_order}), $\omega_{\pm}$ are frequencies of oscillation of the driving fields and
$\Omega_{\pm}^{(1)}$, $\Omega_{\pm}^{(2)}$, $\Omega_{\pm}^{(3)}$ are amplitudes of first-, second- and third-order terms, respectively. As discussed
in appendix \ref{append1} for a simple
two-level system, the rotating wave approximation demands $\Omega_{\pm}^{(3)}\ll\Omega_{\pm}^{(2)}\ll\Omega_{\pm}^{(1)} $; therefore, we choose the amplitudes of the driving field appropriately in our numerical calculations.
The point to be noted is that higher-order terms protect the fluctuations arising from the preceding order term. For example, for a driving field
up to second order, the first-order term evolves the initial state $\vert 0\rangle$ (eigenstate of NV Hamiltonian without driving field)
to dressed states which are protected against fluctuations due to the environment (See the above discussion and appendix for our simple case). However, they are
vulnerable to fluctuations arising due to the driving field itself. The reason for such fluctuations in the driving field is due the
limitation of microwave sources in practical applications.
The second-order term, however, protects these dressed states against the noise in the first-order driving field itself.
%
%

\section{Concatenated Continuous Decoupling
Scheme For NV Spin}
In the preceding section, we discussed a special time-dependent driving field $\Omega(t)$ given by Eq.~(\ref{1st_order}) when considered up to first
order term, Eq.~(\ref{2nd_order}) when a second-order term is also involved and Eq.~(\ref{3rd_order}) when a third-order term is also included.
 This driving is special in the sense it protects the state of
the NV spin against the fluctuations from the environment and the driving field itself.

In this section, we will solve the time-dependent Hamiltonian
Eq.~(\ref{ham}) for this special time-dependent driving using a unitary integration method discussed in Ref.~\cite{rau1,rau2,rau3}.
In order to solve this time-dependent Hamiltonian, let us present it in a block diagonal form \cite{rau1,rau2,rau3} as
\begin{eqnarray}
 \+{H}=\+{H}^{(N)}=\begin{pmatrix}
\+{H}^{(N-n)}& \+{V}	 \\
            \+{V}^{\dagger} & \+{H}^{(n)}
           \end{pmatrix}.
\label{ham_mat}
\end{eqnarray}
In the above equation, the diagonal blocks $H^{(N-n)}$ and $H^{(n)}$ are $N-n$- and $n$- dimensional square matrices.
In what follows, the superscript in parentheses indicates the dimension of the square matrices. Here $N=3$, $n=1$ for our interest:
\begin{eqnarray}
\+{H}^{(2)}&=& \begin{pmatrix}
           H_{11}&H_{12}\\
           H_{21}&H_{22}
          \end{pmatrix},\+{H}^{(1)}=H_{33}, \nonumber \\
\+{V}&=&\begin{pmatrix} H_{13}\\H_{23}\end{pmatrix}, \+{V}^{\dagger}=\begin{pmatrix} H_{31}&H_{32}\end{pmatrix},
\end{eqnarray}
but we set up what follows more generally. $H_{11}, H_{12},\ldots$ are elements of the Hamiltonian.

The time evolution of this Hamiltonian can be written as $\+{U}^{(3)}(t)=\tilde{U}_1\tilde{U}_2$ \cite{rau1,rau2,rau3}, where
\begin{eqnarray}
 \tilde{U}_1=\begin{pmatrix}
              \+{I}^{(2)}&\+{z}(t)\\
                 \+0^{\dagger}&\+{I}^{(1)}
             \end{pmatrix}\begin{pmatrix}
                \+I^{(2)}&\+0\\
                 \+{w}^{\dagger}(t)&\+{I}^{(1)}
                \end{pmatrix},
\end{eqnarray}
and
\begin{eqnarray}
 \tilde{U}_2=\begin{pmatrix}
             \tilde{\+U}^{(2)} & \+0\\
             \+0^{\dagger}&\tilde{\+U}^{(1)}
             \end{pmatrix}.
\end{eqnarray}
Here, $\+{I}^{(2)}$ and $\+{I}^{(1)}$ are identity matrices of dimensions $2$ and $1$, respectively, and
\begin{eqnarray}
 \+{z}(t)=\begin{pmatrix} z_1(t)\\z_2(t)\end{pmatrix},\+{w}^{\dagger}(t)=\begin{pmatrix} w_1(t)&w_2(t)\end{pmatrix},
\end{eqnarray}	
for our case.
$\+0$ is a zero vector of dimension two. $\tilde{\+U}^{(2)}$ and $\tilde{\+U}^{(1)}$
are block diagonals of the evolution operator $\+{U}^{(3)}(t)$, and tilde symbolizes that matrices need not be unitary.
\begin{figure}[!t]
\includegraphics [angle=0,width=.9\columnwidth] {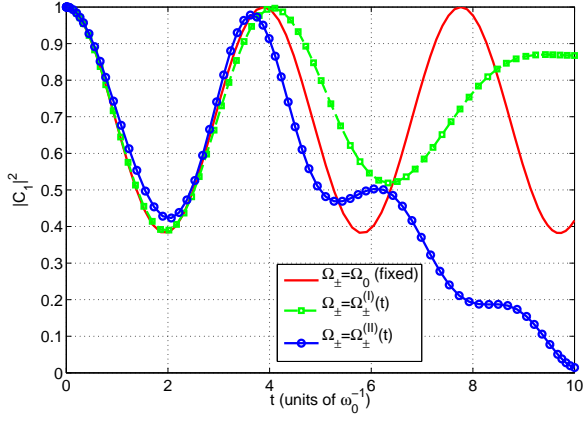}
\caption{(color online) Level population $\vert C_1\vert^2$ plotted as functions of time for a fixed driving, a first-order driving field and
a second-order driving field
given by Eqs. (\ref{1st_order}) and (\ref{2nd_order}).  The parameters for first- and second-order driving fields are
$\omega_+=\omega_-=0.15$, $\Omega_{\pm}^{(1)}=0.9$, $\Omega_{\pm}^{(2)}=\Omega_{\pm}^{(1)}/2$ and $\Delta_{+}=-1$.  Oscillations in level population due to the oscillating
 driving field can be easily distinguished. Time scale is defined by zero field splitting $\omega_{0}$ and is of the order of a nanosecond. }
\label {level_pop1}
\end{figure}
\begin{figure}[!h]
  \includegraphics[width=.9\columnwidth]{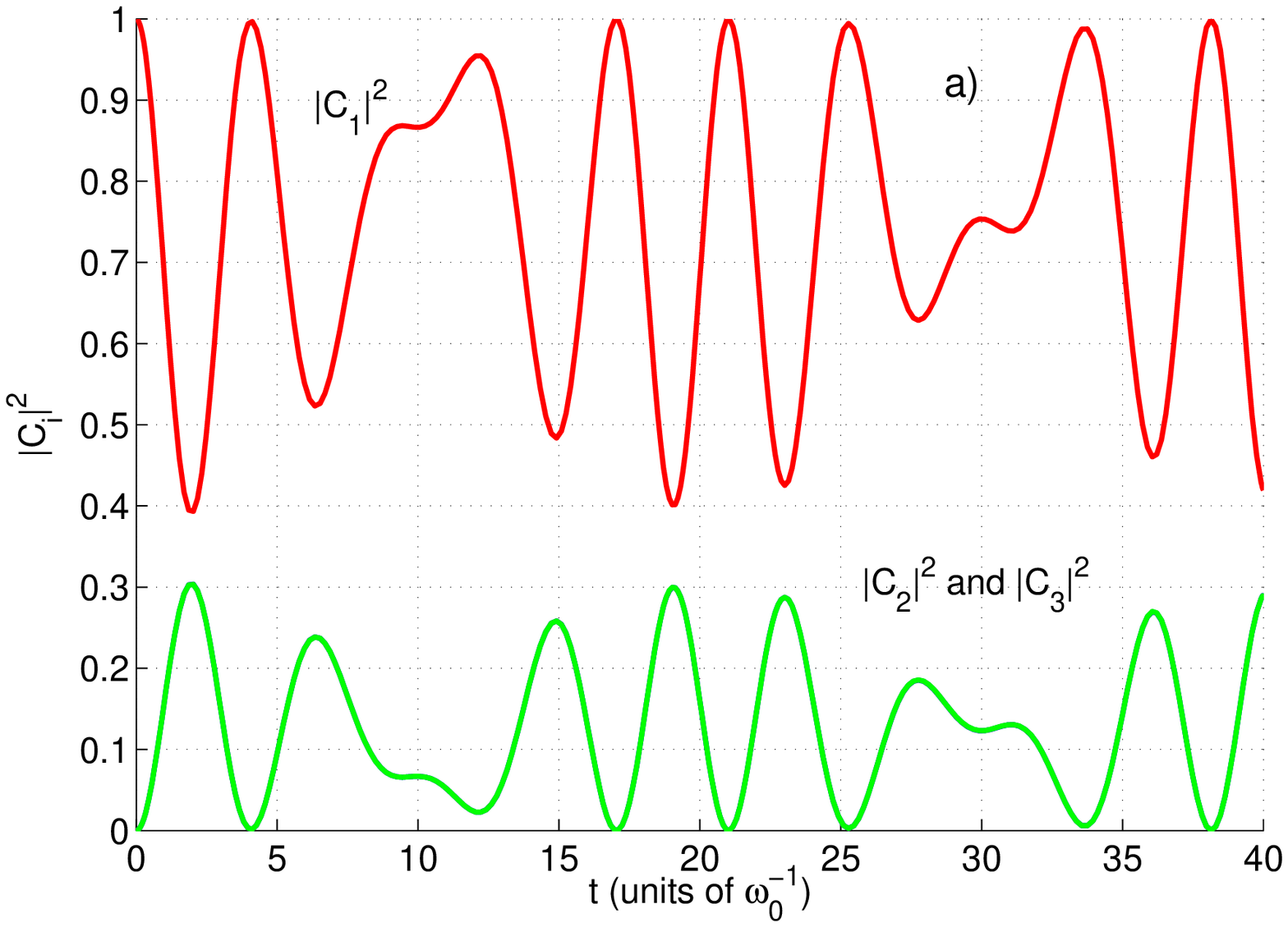} \\
    \includegraphics[width=.9\columnwidth]{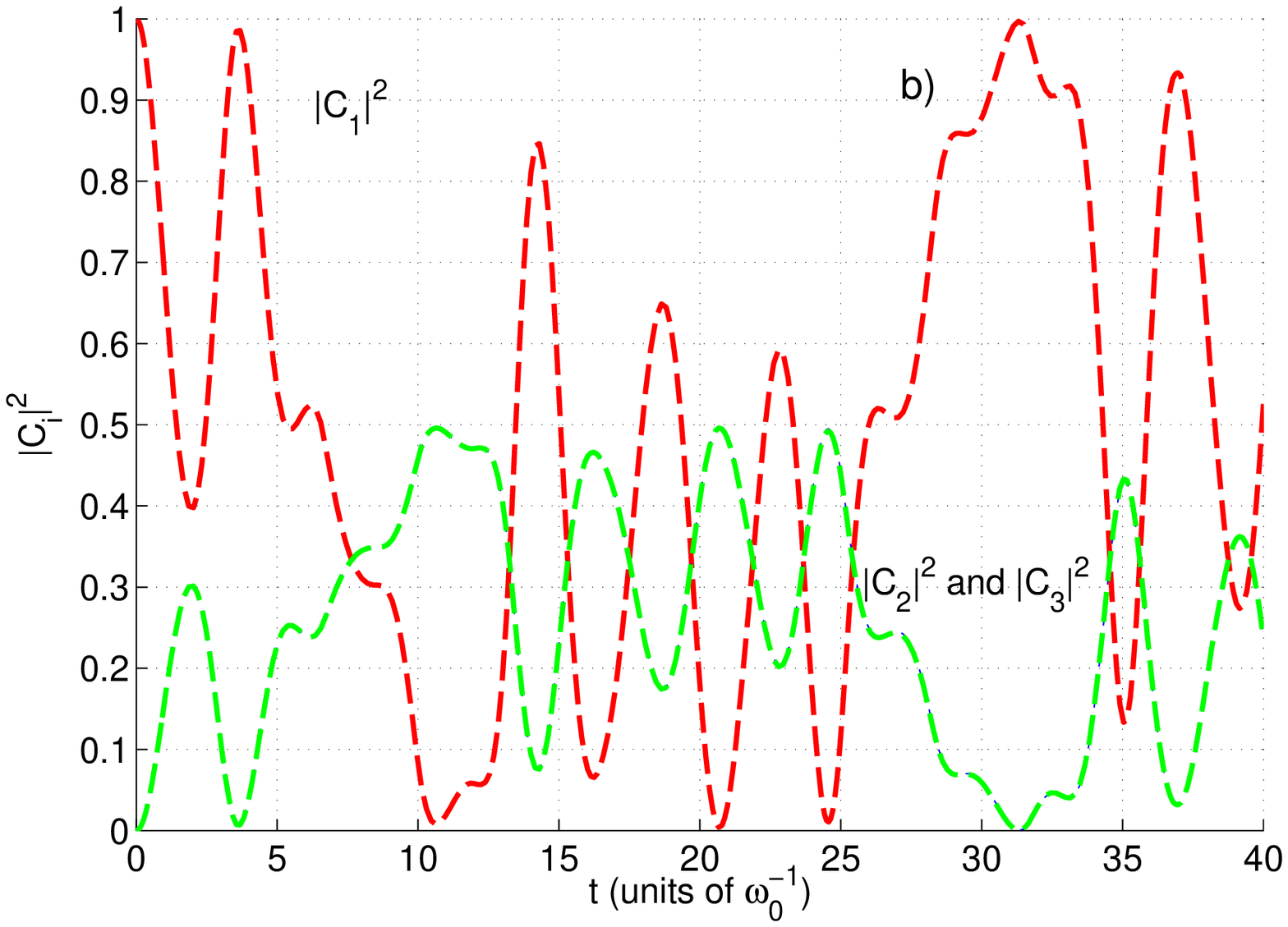}
  \caption{(color online) Level populations $\vert C_i\vert^2=\vert\langle i\vert \psi(t)\rangle\vert^2 $, $i=1, 2, 3$ as functions of time. (a) Solid
lines correspond to first-order and (b) dashed lines to
second-order driving fields. Parameters same as in Fig.~\ref{level_pop1}, $\omega_+=\omega_-=0.15$, $\Omega_{\pm}^{(1)}=0.9$, $\Omega_{\pm}^{(2)}=\Omega_{\pm}^{(1)}/2$ and $\Delta_{+}=-1$,
but results extended to larger time. Green and blue curves of Fig.~\ref{level_pop1} now in red, solid and dashed, respectively. Time scale is defined by zero field splitting $\omega_{0}$ and is of the order of a nanosecond. }
      \label{level_pop}
  \end{figure}
\begin{figure}[!h]
\includegraphics [angle=0,width=.9\columnwidth] {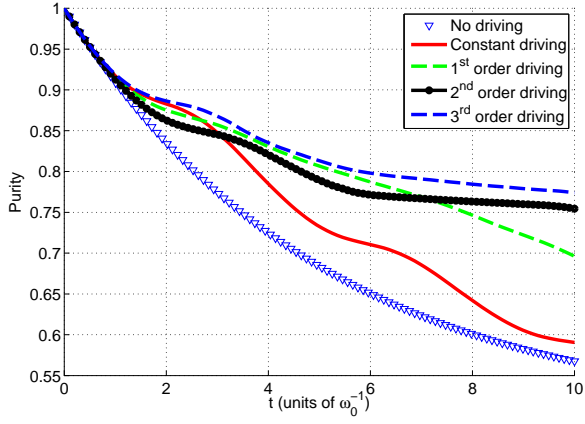}
\caption{ (color online) Purity is plotted as a function of time for constant, first-order and second-order driving fields. For comparison, the case of no driving field is also shown. As we include higher order driving fields,
the purity of the system improves significantly. In all the cases, $\Gamma=0.05$ and $\Delta_{+}=0.9$.
For time-dependent driving fields, the parameters are: $\omega_+=1.0$, $\omega_-=0.35$, $\Omega_{+}^{(1)}=1.0$, $\Omega_{-}^{(1)}=0.8$,
$\Omega_{\pm}^{(2)}=\Omega_{\pm}^{(1)}/2$ and $\Omega_{\pm}^{(3)}=\Omega_{\pm}^{(2)}/2$. Time scale is defined by zero field splitting $\omega_{0}$ and is of the order of a nanosecond.}
\label {order_field}
\end{figure}

The block diagonal form of
the unitary operator is advantageous to solve the equation of motion $i\dot{\+U}^{(3)}={\+H}^{(3)}(t){\+U}^{(3)}(t)$, which reduces to (an over-dot will denote derivative with respect to time)
\begin{eqnarray}
 i\dot{\tilde{U}}_2=\+H^{\rm eff}(t)\tilde{U}_2,
\label{u2eq}
\end{eqnarray}
where
\begin{eqnarray}
 \+H^{\rm eff}=\tilde{U}_1^{-1}\+H^{(3)}\tilde{U}_1-i\tilde{U}_1^{-1}\dot{\tilde{U}}_1.
\end{eqnarray}
As $\tilde{U}_2$ is block diagonal, the right-hand side of Eq.~(\ref{u2eq}) must ensure the vanishing of the off-diagonal blocks of $\+H_{\rm eff}$. This leads to
a condition for $\+z(t)$ as
\begin{eqnarray}
 i\dot{\+z}=\+{H}^{(2)}\+{z}(t)+\+V-\+{z}(t)(\+{V}^{\dagger}\+{z}(t)+\+{H}^{(1)}).
\label{zeq}
\end{eqnarray}

A similar set of equations for $\+{w}(t)$ can also be given, which is related to $\+{z}(t)$ as \cite{rau1,rau2,rau3}
\begin{eqnarray}
 \+z(t)=-\gamma_1\+w(t)=-\+w(t)\gamma_2,
\end{eqnarray}
where $\+{\gamma}_1=\+{I}^{(2)}+\+z(t)\+z^{\dagger}(t)$ and $\+\gamma_2=(\+I^{(1)}+\+z^{\dagger}(t)\+z(t))^{-1}$.
In our particular case, the components of the $\+z(t)$ vector are given by
\begin{eqnarray}
 i\dot{z}_1&=&\frac{\Omega_{-}(t)}{2}z_2(t)+\frac{\Omega_{+}(t)}{2}\bigl(1-\vert z_1(t)\vert^2\bigr)+z_1(t)\Delta_{+},\nonumber \\
i\dot{z}_2&=&\biggl( \frac{\Omega_{-}(t)}{2}-\frac{\Omega_{+}(t)}{2}z_2(t)\biggr) z_1(t).
\label{zcomp}
\end{eqnarray}

Eq.~(\ref{u2eq}), after using Eq.~(\ref{zeq}), has only diagonal blocks on both sides. The components in explicit matrix form are given as
\begin{equation}
\begin{split}
 i\begin{pmatrix}
(\dot{\tilde{U}}_2)_{11} & (\dot{\tilde{U}}_2)_{12} & 0\\
(\dot{\tilde{U}}_2)_{21} & (\dot{\tilde{U}}_2)_{22} & 0 \\
0 & 0 &(\dot{\tilde{U}}_2)_{33}
 \end{pmatrix}&=\begin{pmatrix} H_{11}^{\rm eff} & H_{12}^{\rm eff} & 0\\
                              H_{21}^{\rm eff}& H_{22}^{\rm eff} & 0 \\
                              0& 0 & H_{33}^{\rm eff} \end{pmatrix} \nonumber \\
&\times\begin{pmatrix} (\tilde{U}_2)_{11} & (\tilde{U}_2)_{12} & 0 \\
                                                                                 (\tilde{U}_2)_{21} &(\tilde{U}_2)_{22} & 0 \\
                                                                                 0 & 0 & (\tilde{U}_2)_{33} \end{pmatrix},
\label{explicit}
\end{split}
\end{equation}
where the components of the effective Hamiltonian are
\begin{eqnarray*}
  H_{11}^{\rm eff}&=&-\frac{\Omega_{+}(t)}{2} \biggl( z_1(t)-w_1(t)(1-
\vert z_1(t)\vert^2)\biggr)\nonumber \\
&+&w_1(t)\biggl(\Delta_{+}z_1(t)+\frac{\Omega_{-}(t)}{2}z_2-i\dot{z}_1 \biggr) , \nonumber \\
H_{12}^{\rm eff}&=&\frac{\Omega_{-}(t)}{2}(1+w_2(t)z_2(t)) \nonumber \\
&+&w_2(t)\biggl(
\Delta_{+}z_1(t)+\frac{\Omega_{+}(t)}{2}(1-\vert z_1(t)\vert^2)-i\dot{z}_1 \biggr),\nonumber \\
H_{21}^{\rm eff}&=&(1+w_1(t)z_1(t))\biggl( \frac{\Omega_{-}(t)}{2} -
\frac{\Omega_{+}(t)}{2} z_2(t)\biggr) \nonumber \\
&-&iw_1(t)\dot{z}_2, \nonumber\\
H_{22}^{\rm eff}&=&-\Delta_{+}+w_2(t)\biggl[ z_1\biggl(  \frac{\Omega_{-}(t)}{2}
-\frac{\Omega_{+}(t)}{2}z_2(t)\biggr) -i\dot{z}_2\biggr],
\end{eqnarray*}
and,
\begin{eqnarray}
H_{33}^{\rm eff}&=&-\Delta_{+}\bigl(1+w_1(t)z_1(t)\bigr)- \frac{\Omega_{-}(t)}{2}\bigl(w_2(t)z_1(t) \nonumber \\
&+& w_1(t)z_2(t)\bigr)-\frac{\Omega_{+}(t)}{2}
\bigl( w_1(t)(\vert z_1(t)\vert^2-1)\nonumber \\
&-&z_1(t)(1+w_2(t)z_2(t))\bigl)+i(w_1(t)\dot{z}_1+w_2(t)\dot{z}_2).\nonumber \\
\label{u2comp}
\end{eqnarray}
The components of effective Hamiltonian given by Eq.~(\ref{u2comp}), together with Eq.~(\ref{zcomp}), is sufficient to calculate the
operator $\tilde{U}_2$.
Also the solutions for $\+z(t)$ (and $\+w(t)$) are enough to calculate $\tilde{U}_1$.
Here we note that operators $\tilde{U}_1$ and $\tilde{U}_2$ are not unitary.
A unitarization procedure can be performed by following Ref.~\cite{rau2} although it is not necessary, their product guaranteed to be unitary for a Hermitian Hamiltonian by the above construction. For the sake of unitarization, we should
take into account the product
\begin{eqnarray}
 \tilde{U}_1^{\dagger}\tilde{U}_1=\begin{pmatrix}
                                   \+\gamma_1^{-1} & 0 \\
                                        0^{\dagger} & \+\gamma_2
                                  \end{pmatrix}=\begin{pmatrix}	\+g_1\+g_1^{\dagger} & 0\\
                                                   0^{\dagger}& \+g_2\+g_2^{\dagger} \end{pmatrix}^{-1},
\end{eqnarray}
and perform a gauge transformation using Hermitian square root matrices $\+g_1$ and $\+g_2$ as
\begin{eqnarray}
 U_1=\tilde{U}_1\begin{pmatrix}\+{g_1} & 0 \\
                 0^{\dagger} & \+{g_2}
                \end{pmatrix},
\end{eqnarray}
and
\begin{eqnarray}
 U_2=\begin{pmatrix}\+{g_1}^{-1} & 0 \\
                 0^{\dagger} & \+{g_2^{-1}}
                \end{pmatrix}\tilde{U}_2.
\end{eqnarray}

A useful quantity for investigation is the level population, defined as $\vert C_i\vert^2=\vert\langle i\vert \psi(t)\rangle\vert^2 $, where
$i=1$, $2$, $3$ and $\psi(t)=U^{(3)}(t)\vert1\rangle$. Here levels $\vert0\rangle$, $\vert-1\rangle$ and  $\vert1\rangle$ in Fig.~\ref{three_lev}
 are relabeled as $\vert1\rangle$, $\vert2\rangle$ and  $\vert3\rangle$. In all the studies below, we calculate the time evolution
from $t=0$ to $t_{\max}$.
 Fig.~\ref{level_pop1} displays  the population in the first level as a function of time
for fixed driving and oscillating driving fields given by Eq.~(\ref{1st_order}) and Eq.~(\ref{2nd_order}). The parameters are set arbitrarily.
 The oscillations in the driving field
change the level population significantly.
In Fig. \ref{level_pop}, we extend to larger time, again with arbitrary set of parameters, and show also
populations for all three levels. The
evolution operator is calculated by the method discussed above. In figures \ref{level_pop1} and \ref{level_pop}, since we
consider $\Omega_{+}(t)=\Omega_{-}(t)$, this results in equal population in level 2 and 3, so that
 $\vert C_2^2\vert$ and $\vert C_3^2\vert$ overlap. For the first-order driving, we see that as time progresses the
population in level 1 decreases from its maximum value, reaching a lowest value of $0.4$. Correspondingly, population in
levels 2 and 3 rises from zero to $0.3$ for both. For longer times, level populations oscillate, with recurrences of the populations at $t=0$.
For the case of second-order driving, similar oscillations and recurrences occur albeit at different values of $t$, with the additional feature that the
population in level 1 drops to zero at certain points, levels 2 and 3 correspondingly rising to $0.5$.
\begin{figure}[!t]
\includegraphics [angle=0,width=.85\columnwidth] {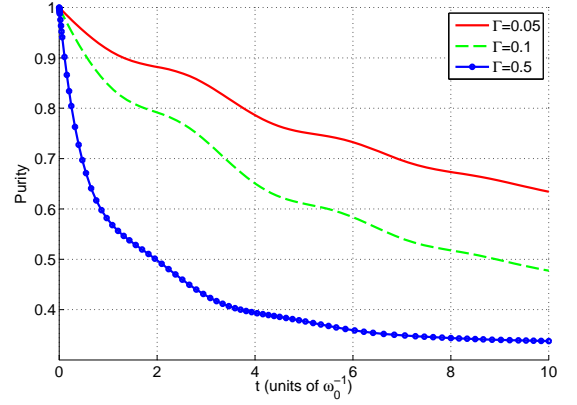}
\caption{ (color online) Purity for second-order driving field is plotted as a function of time for $\Gamma=0.05$, $0.1$ and $0.5$. As in previous figures,
$\omega=0.15$, $\Omega_{\pm}^{(1)}=0.9$, $\Omega_{\pm}^{(2)}=\Omega_{\pm}^{(1)}/2$ and $\Delta_{+}=-1$. It can be seen that increasing the decoherence $\Gamma$ reduces the purity. Time scale is defined by zero field splitting $\omega_{0}$ and is of the order nanosecond.}
\label {purity_cas}
\end{figure}

Let us expand the discussion to systems with dissipation. It should be noted that till now we have not included
fluctuations due to environment or driving field. We first analyze the system using Liouville-von Neumann-Lindblad
equation containing dissipation and decoherence term and solve it using the unitary integration method. We will start with Liouville-von Neumann-Lindblad
 master equation for the density
matrix, which has the following form \cite{lindblad,sudarshan}:
\begin{eqnarray}
 i\dot{\rho}=[\+{H},\rho]-\frac{i}{2}(\+L^{\dagger}\+L\rho +\rho \+L^{\dagger} L-2 L \rho \+L^{\dagger}).
\label{lindb}
\end{eqnarray}
with $\+{H}$ given by Eq.~(\ref{ham_mat}).
The Lindblad opeartor $\+L$ in the above equation introduces irreversible dissipation and decoherence to the system and is taken in the form:
\begin{eqnarray}
 L=\sqrt{\Gamma} \lambda_i^{(G)}=\sqrt{\Gamma}\begin{pmatrix}
                                               0&1&0\\
                                               1&0&0\\
                                               0&0&0
                                              \end{pmatrix}.
\end{eqnarray}
Here $\lambda_i^{(G)}$ are the Gell-Mann matrices \cite{joshi}.

$\Gamma$ in the above equation sets the rate of relaxation
due to dissipation. It may be noted that the evolution
of $\rho$ in Eq.~(\ref{lindb}) may be nonunitary but the form of the equation preserves trace operation of $\rho$ as well as positivity of probabilities.
More mathematical details related to such superoperators and dynamical semigroups can be referred to \cite{alicki}.
Substituting the above form of the Lindblad operator into Eq.~(\ref{lindb}), the nine elements of the
density matrix can be written as
\begin{eqnarray}
 i\dot{\rho}_{11}&=&-i\Gamma(\rho_{11}-\rho_{22})-\frac{\Omega_{-}(t)}{2}
\rho_{12}-\frac{\Omega_{+}(t)}{2}\rho_{13} \nonumber \\
&+&\frac{\Omega_{+}(t)}{2}\rho_{31}+\frac{\Omega_{-}(t)}{2}\rho_{21}, \nonumber \\
 i\dot{\rho}_{12}&=& i\Gamma\rho_{21}+(\Delta_{+}-i\Gamma)\rho_{12}+\frac{\Omega_{-}(t)}{2}(\rho_{22}-
\rho_{11})\nonumber \\
&+&\frac{\Omega_{+}(t)}{2}\rho_{32},\nonumber \\
 i\dot{\rho}_{13}&=&\frac{\Omega_{+}(t)}{2}(\rho_{33}-\rho_{11})+(\Delta_{+}-i\frac{\Gamma}{2})\rho_{13}+
\frac{\Omega_{-}(t)}{2}\rho_{23},\nonumber \\
 i\dot{\rho}_{21}&=& i\Gamma\rho_{12}-(\Delta_{+}+i\Gamma)\rho_{21}+\frac{\Omega_{-}(t)}{2}(\rho_{11}-
\rho_{22})\nonumber \\
&-&\frac{\Omega_{+}(t)}{2}\rho_{23},\nonumber \\
 i\dot{\rho}_{22}&=&\frac{\Omega_{-}(t)}{2}(\rho_{12}-\rho_{21})+i\Gamma(\rho_{11}-\rho_{22}), \nonumber \\
 i\dot{\rho}_{23}&=&\frac{\Omega_{-}(t)}{2}\rho_{13}-\frac{\Omega_{+}(t)}{2}\rho_{21}-i\frac{\Gamma}{2}\rho_{23},\nonumber \\
 i\dot{\rho}_{31}&=&\frac{\Omega_{+}(t)}{2}(\rho_{11}-\rho_{33})-(\Delta_{+}+i\frac{\Gamma}{2})\rho_{31}
-\frac{\Omega_{-}(t)}{2}\rho_{23},\nonumber \\
 i\dot{\rho}_{32}&=&-\frac{\Omega_{-}(t)}{2}\rho_{31}+\frac{\Omega_{+}(t)}{2}\rho_{12}-i\frac{\Gamma}{2}\rho_{32},\nonumber \\
i\dot{\rho}_{33}&=&\frac{\Omega_{+}(t)}{2}(\rho_{13}-\rho_{31}).\nonumber \\
\label{rho_element}
\end{eqnarray}

Though we have shown explicitly the evolution of the nine components of the density matrix in terms of nine
coupled equations, the tracelessness of the right hand side of Eq.~(\ref{lindb}) guarantees the preservation of $\rm{Tr}\rho$.
Hence one equation can be eliminated during the evolution. 
In order to calculate the evolution of the density matrix, we define
$\xi_1=\rho_{11},\ \xi_2=\rho_{12},\ \xi_3=\rho_{13}, \ \xi_4=\rho_{21},\
\xi_5=\rho_{22},\  \xi_6=\rho_{23},\  \xi_7=\rho_{31}, \ \xi_8=\rho_{32} $. 
Hence, instead of the Hamiltonian (Eq.~\ref{ham}) and Liouville-von Neumann-Lindblad  equation of motion Eq.~(\ref{lindb}), we can now cast it in the form
\begin{eqnarray}
 i\dot{\xi}=\mathcal{P}(t)\xi(t),
\label{xieqn}
\end{eqnarray}
\begin{figure}[!t]
\includegraphics [angle=0,width=.8\columnwidth] {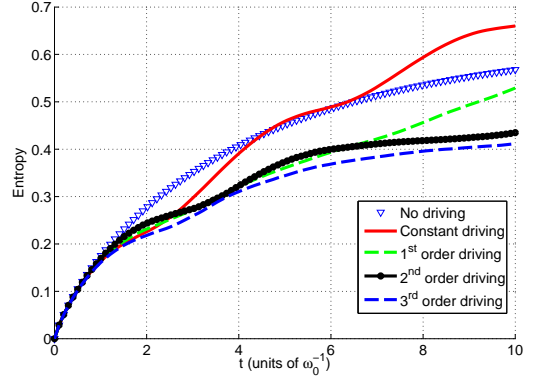}
\caption{(color online) The evolution of entropy is plotted for no driving, constant driving, first-order and second-order drivings.
In all the cases, $\Gamma=0.05$ and $\Delta_{+}=0.9$.
For time-dependent driving fields, the parameters are: $\omega_+=1.0$, $\omega_-=0.35$, $\Omega_{+}^{(1)}=1.0$, $\Omega_{-}^{(1)}=0.8$,
$\Omega_{\pm}^{(2)}=\Omega_{\pm}^{(1)}/2$ and $\Omega_{\pm}^{(3)}=\Omega_{\pm}^{(2)}/2$. Time scale is defined by zero field splitting $\omega_{0}$ and is of the order of a nanosecond.}
\label {entro_cal}
\end{figure}
where
\begin{eqnarray}
 \dot{\xi}(t)=\biggl(
               \dot{\xi}_1(t), \ \dot{\xi}_2(t),\ \dot{\xi}_3(t), \ \dot{\xi}_4(t), \ \dot{\xi}_5(t), \nonumber \\
 \dot{\xi}_6(t), \ \dot{\xi}_7(t), \ \dot{\xi}_8(t), \ \dot{\xi}_9(t)
              \biggr).
\end{eqnarray}
and $\mathcal{P}(t)$ is a $9\times9$ matrix whose elements are drawn from Eq.~(\ref{rho_element}).
We solve Eq.~(\ref{xieqn}) for the particular type of time-dependent driving fields discussed
in Eqs. (\ref{1st_order}) and (\ref{2nd_order}). A quantity of
interest is purity that quantifies entanglement between driving field and atom. It is given as
\begin{eqnarray}
 P(t,\vert\psi\rangle)=\rm{Tr}[\hat{\rho}^2],
\end{eqnarray}
where $\hat{\rho}=\vert\psi(t)\rangle\langle\psi(t)\vert$, and

\begin{eqnarray}
 \vert\psi(t)\rangle=C_1(t)\vert 0\rangle+C_2(t)\vert-1\rangle+C_3(t)\vert 1\rangle.
\end{eqnarray}

In Fig.~\ref{order_field}, the purity $P(t,\vert\psi\rangle)$ is plotted as a function of time
for no driving, constant driving, first-order, second-order and third-order driving fields. The first-, second- and third- order drivings are
given by Eqs. (\ref{1st_order}), (\ref{2nd_order}) and (\ref{3rd_order}). In all the cases, $\omega_+=1.0$ and $\omega_-=0.35$. It can be seen that higher-order
fields improve the purity of the state. In order to maximize the purity in our numerical calculation we set
the amplitude of the successive order terms as $\Omega_{\pm}^{(2)}=\Omega_{\pm}^{(1)}/2$ and $\Omega_{\pm}^{(3)}=\Omega_{\pm}^{(2)}/2$.
Along with these amplitudes, other optimized parameters are $\Omega_{+}^{(1)}=1.0$, $\Omega_{-}^{(1)}=0.8$ $\omega_+=1.0$, $\omega_-=0.35$ and $\Delta_+=0.9$. We illustrated the dependence on the parameter
$\Gamma$ of the purity in Fig.~\ref{purity_cas} for driving fields up to second-order terms for an arbitrary choice of parameters. We see that the
smaller the value of $\Gamma$, the more pure the state of the system at any particular time.

In Fig.~\ref{entro_cal}, we calculate the time evolution of another quantity besides purity, namely,
entropy defined as $S=-\rm{Tr}(\rho \log_3\rho)$ \cite{rau4}.  For this three-level system, we use the base-3 logarithm, so as to standardize entropy
to be zero for a pure state and unity for the opposite case of a completely
random mixed state.  The evolution of the entropy towards an asymptotic value is consistent with the purity evolution shown in Fig.~\ref{order_field}.
 With any dissipation, coherences are finally lost, the density matrix reaching a diagonal form. For the populations noted earlier before dissipation
is introduced, taking those sets as the entries of a diagonal density matrix predicts values of $P$ and $S$ as
follows: for $(0,1/2,1/2), P=1/2, S=0.631$, for $(0.4,0.3,0.3), P=0.34, S=0.991,$ and for the
completely random state $(1/3,1/3,1/3), P=1/3, S=1$. Note that the values of purity and entropy nearly coincide for the last two cases.

\section{Fluctuations due to spin bath}

With their unique physical properties, NV centers are attractive for application in solid state quantum information processing.
However, considering NV centers as isolated objects and neglecting environmental effects is unrealistic. Most NV centers are surrounded by $^{13}$C nuclear spins and, therefore, cannot be completely decoupled from the unwanted interactions with a $^{13}$C spin-based environment. The loss of  coherence of the central NV spin by a spin bath of surrounding $^{13}$C nuclei has been a serious concern
for the usage of NV centers in quantum computing.

In the previous sections, we demonstrated the advantage of the CCD driving protocol in minimizing  decoherence effects in the three-level NV spin model.
However, that treatment of decoherence effects was quite general. In this last concluding section, we will study particular aspects of the CCD driving
protocol for an NV center coupled to the $^{13}$C spin bath. This coupling leads to a random fluctuating contribution in the
detuning term $\Delta_{i}$ in Eq.~(\ref{ham}) that cannot be completely captured by a purely dissipative term in the Lindblad master equation.
In what follows, we demonstrate that the CCD driving protocol extends also to the random fluctuations and thus
 can minimize decoherence effects that are also specific to the $^{13}$C spin bath. In doing so we will also take care of
fluctuations arising from the microwave source itself.

The hyperfine coupling of the NV spin to the $^{13}$C nuclear spins causes a dephasing of the
central spin \cite{9,17,ryan,xu,huang,zhao}. In principle, the quantum state experiences a random field due the
bath spins. Such effects are generic in nature and also applicable for other solid state qubits like spins in quantum dots.
The effect of the fluctuating random field term can be represented by a
term $\mathcal{H}_{\rm hf}=S_z\hat{z}.\sum_j{\bf A}_j.\vec{I}_j (\equiv S_zb_z)$ in
the Hamiltonian, where ${\bf A}_j$ is the hyperfine coupling of the $j^{\rm th}$ nuclear spin $\vec{I}_j$ to the NV spin and $b_z=\sum_j{\bf A}_j.\vec{I}_j$.
Here we neglect
the transverse component of the hyperfine coupling because the effects
arising from that component are negligibly small \cite{huang}. This hyperfine coupling
provides effectively a fluctuating field $b_z$.

In a mean field approximation, the random field $b_z$ can be considered as a mean field
due to all the neighboring $^{13}$C nuclei that results in a net fluctuating detuning of the driving frequency of the pulse from resonance \cite{delange}.
Thus the random field $b_z$ can be incorporated in the Hamiltonian of the NV spin by considering an effective detuning $\Delta_+ +\zeta(t)$ and
$\Delta_-+\zeta(t)$ to the excited levels $m_S=+1$ and $m_S=-1$, where $\zeta(t)$ is a random time-dependent sequence incorporating the fluctuating field.
The modified Hamiltonian of the NV spin in the presence of random fields can thereby be written as
\begin{eqnarray}
 H_{NV}=\sum_{i=\pm1}\biggl((-\Delta_i+\zeta(t)) \vert i\rangle\langle i \vert+\frac{\Omega_i}{2}\bigl(\vert 0 \rangle\langle i
\vert + (\vert i \rangle\langle 0 \vert \bigr)\biggr). \nonumber \\
\label{ham_new}
\end{eqnarray}
\begin{figure}[!t]
  \includegraphics[width=.9\columnwidth]{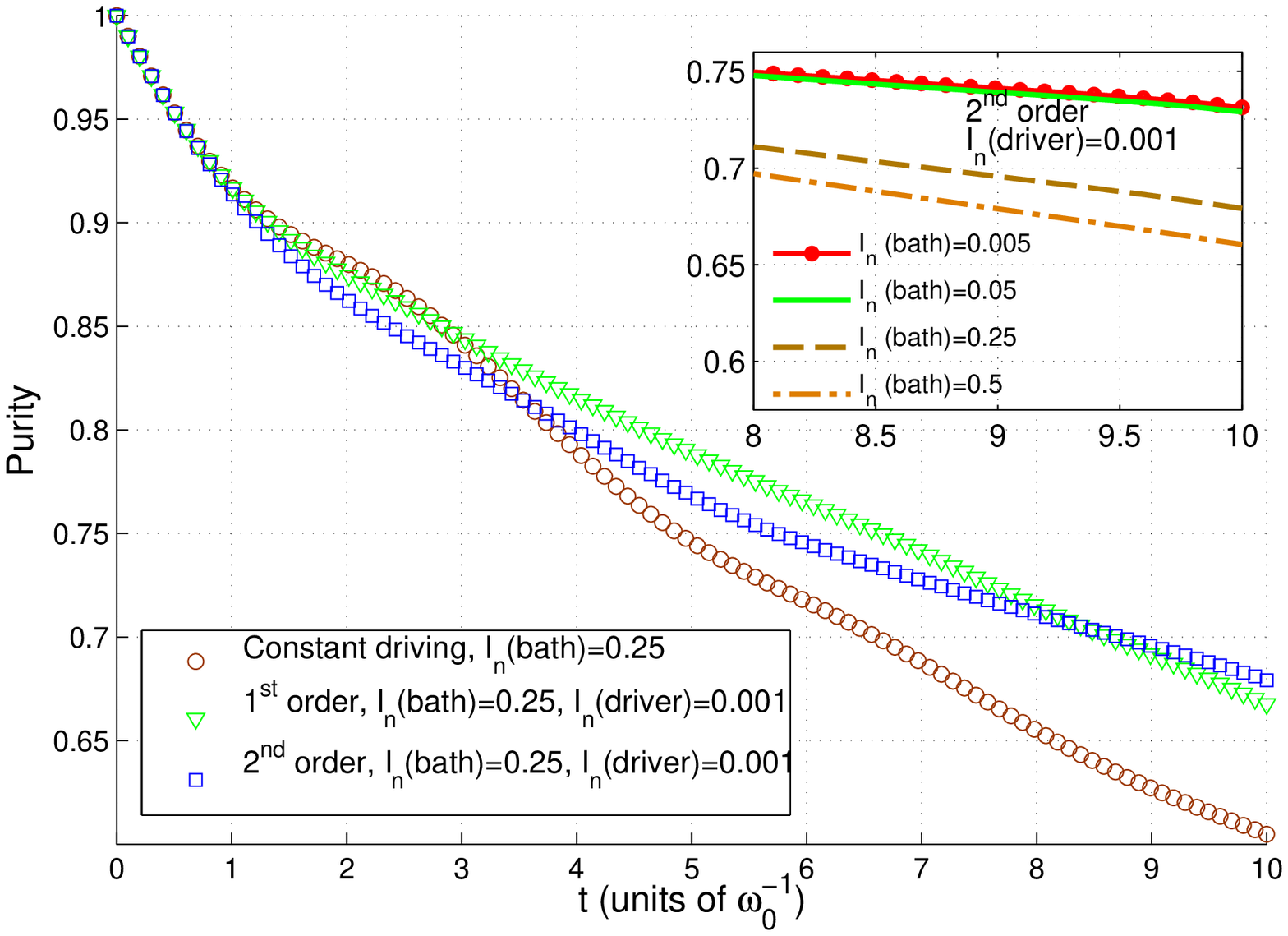} \\
    \includegraphics[width=.9\columnwidth]{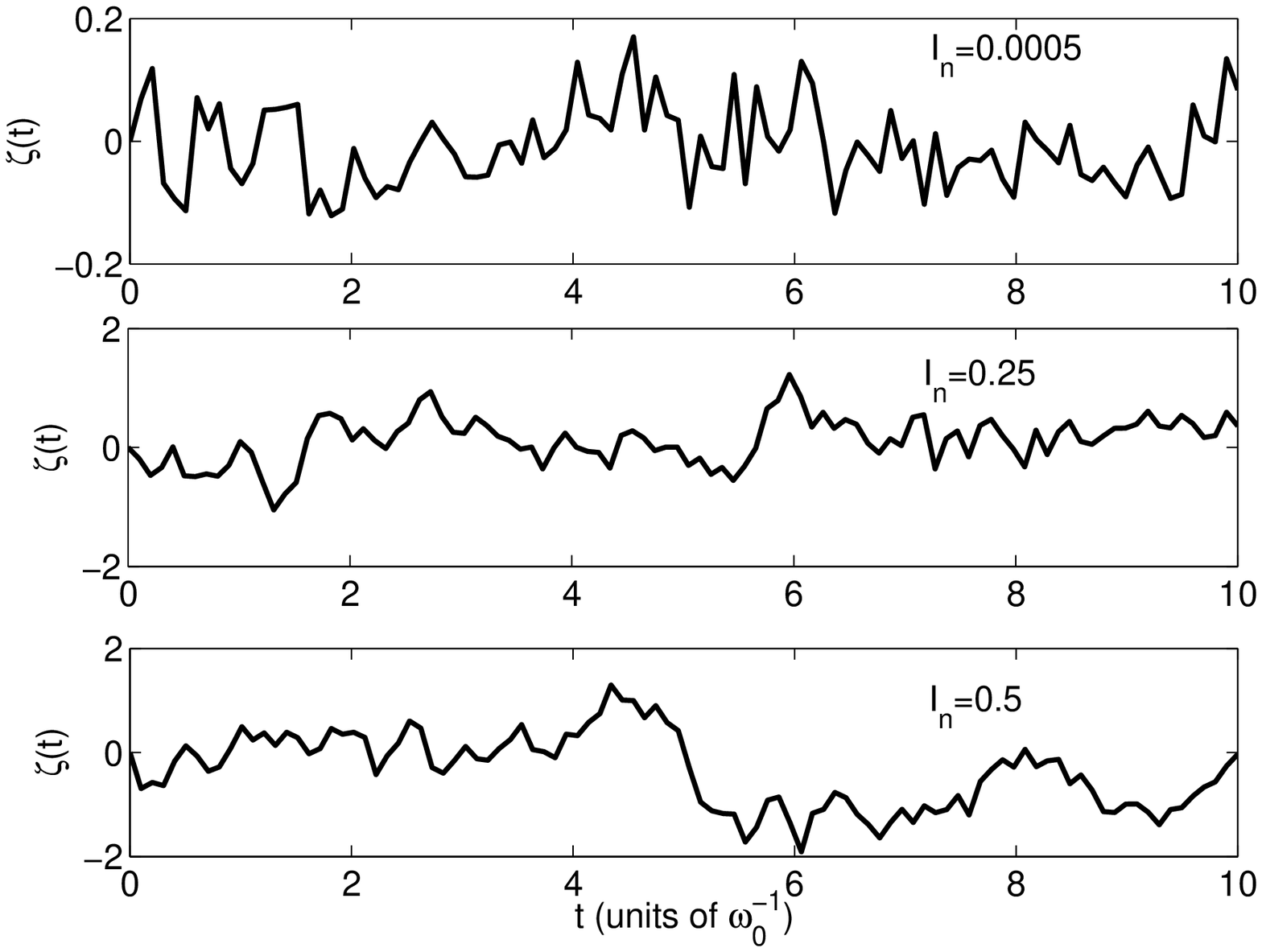}
  \caption{(color online) Purity in the presence of noise of the surrounding
heat bath for a constant driving; for first-order driving with noise from the MW source itself;
and second-order driving case, when the second order term overcomes
 the fluctuation in $\Omega^{(1)}_{\pm}$. The inset shows second-order driving for varying noise intensities of the surrounding spin bath with a
fixed noise ($I_n=0.001$) in the amplitude $\Omega^{(1)}_{\pm}$. Periodic drivings improve the purity
in the presence of noise as compared to the constant driving case. Increasing the noise strength lowers the purity, as expected. The various noises for different intensities are also shown. The numerical simulation has been done for 1000 realizations.
 Parameters are the same as in Fig.~\ref{order_field}, $\omega_+=1.0$, $\omega_-=0.35$, $\Omega_{+}^{(1)}=1.0$, $\Omega_{-}^{(1)}=0.8$,
 $\Omega_{\pm}^{(2)}=\Omega_{\pm}^{(1)}/2$ and $\Delta_{+}=0.9$. The
relaxation parameter is taken as $\Gamma=0.05$.
Time scale is defined by zero field splitting $\omega_{0}$ and is of the order of a nanosecond. }
      \label{with_fluc}
  \end{figure}
Now we investigate the response of CCD driving in the presence of fluctuating random fields by evolving the Hamiltonian (\ref{ham_new})
and calculating again the purity and entropy with this new Hamiltonian. Our aim is to demonstrate that the CCD driving protocol is still efficient, even
in the presence of noise due to surrounding spin bath and a noise due to the driving field source itself.
 The random field term in the above Hamiltonian as well as in the driving field can me modeled as an Ornstein-Uhlenbeck (OU)
process governed by \cite{dt}
\begin{eqnarray}
 \zeta(t+ dt)=\zeta(t)e^{-dt/\tau}+\sqrt{\frac{c\tau}{2}\biggl(1-e^{-2dt/\tau}\biggr)} n_r,
\end{eqnarray}
where $n_r$ is a unit normal random variable with zero mean and unit variance, and $\zeta(0)=0$. $\tau$ and $c$ are positive constants called
relaxation time and diffusion constant, respectively, of the
$^{13}$C nuclear spins in the vicinity of the NV spin of interest. The variance of the OU process is defined as $\sigma^2=c\tau/2$ and the autocorrelation
function is given by $\langle\zeta(0)\zeta(t)\rangle=\sigma^2e^{-|t|/\tau}$ \cite{hanggi}. $I_n=\sigma^2\tau$ ($\equiv c\tau^2$) is a measure of the intensity of noise.
\begin{figure}[!t]
\includegraphics [angle=0,width=.8\columnwidth] {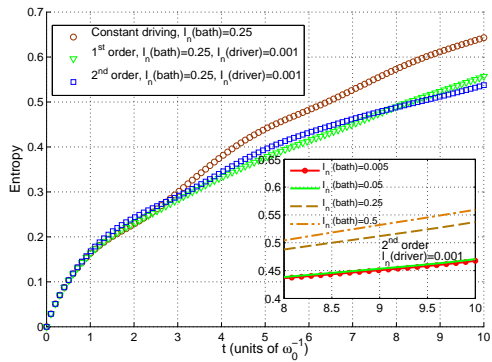}
\caption{(color online) The evolution of entropy is plotted in the presence of noise from the surrounding spin bath
for a constant driving; for first-order driving with noise from the MW source itself;
and second-order driving case, when the second order term overcomes
 the fluctuation in $\Omega^{(1)}_{\pm}$.
The inset shows second-order driving for varying noise intensities of the surrounding spin bath with a
fixed noise ($I_n=0.001$) in the amplitude $\Omega^{(1)}_{\pm}$.
Parameters are the same as in Fig.~\ref{entro_cal}, $\omega_+=1.0$, $\omega_-=0.35$, $\Omega_{+}^{(1)}=1.0$, $\Omega_{-}^{(1)}=0.8$,
 $\Omega_{\pm}^{(2)}=\Omega_{\pm}^{(1)}/2$ and $\Delta_{+}=0.9$ and the
relaxation parameter $\Gamma=0.05$.
Time scale is defined by zero field splitting $\omega_{0}$ and is of the order of a nanosecond.}
\label {entro_fluc}
\end{figure}

In our numerical calculation, the amplitude of random field fluctuations can be
varied by changing the noise intensity $I_n$. We investigated the time evolution of purity using the density matrix given
by Eq.~(\ref{lindb}) for the Hamiltonian in Eq.~(\ref{ham_new}) and relaxation parameter $\Gamma=0.05$. Fig.~\ref{with_fluc}
shows the purity for various cases with constant and periodic drivings in the presence of noise with intensity $I_n=0.25$ due to spin bath of
surrounding $^{13}$C nuclear spins.
In order to improve the purity of the system when a first-order periodic driving is applied, the fluctuations inherent within the
driving field start dephasing the
quantum state of the NV spin. Here we consider the noise due to the MW source very small ($I_n=0.001$) in comparison to the noise due to the surrounding spins.
The second-order term as discussed in the previous section and given by Eq.~(\ref{2nd_order}) overcomes the fluctuations arising due to the first order term
and improves the purity.
The averaging is done over 1000 realizations. 
The inset of Fig.~\ref{with_fluc} shows the effect of noise on the purity for the second-order CCD driving case where fluctuations are arising due to the
neighboring spin bath and due to the MW source itself. The fluctuations due to the MW source is manifested in the amplitude of the first-order driving
field. As the noise strength increases,
the purity of the system declines but not as much as in the case of constant driving.
Hence, CCD driving
holds merit over constant driving and dephasing is minimized in the presence of the inherent noise of driving fields.

Similar analysis for entropy is shown in Fig.~\ref{entro_fluc}.  Here we see that the entropy for the CCD driving
case is smaller than in the case of constant driving. However, while overcoming the fluctuations due to the neighboring spins
and the source itself, the introduction of higher order driving fields lowers the entropy. The findings in Fig.~\ref{entro_fluc}
are consistent with that in Fig.~\ref{with_fluc}. The inset of Fig.~\ref{entro_fluc} shows the evolution of entropy
for second-order driving for different noise intensities in the presence of inherent noise of the driving fields. As expected, again,
 increasing the noise strength increases the entropy. Thus we
conclude that the non-monochromatic CCD driving is useful to minimize losses of the the quantum state of the NV spin in the presence of noise due
to $^{13}$C nuclei, of interest for using NV spins as qubits in quantum information.

\section{Conclusion}
We have studied the NV three-level system by applying  a driving field using a concatenated continuous decoupling (CCD) scheme. A comparative study
for various orders of driving field is presented. In order to study qualitatively the characteristics of the system, we have calculated level
populations, purity, and entropy as signifying entanglement between the field and the atom. Various modes of oscillations in the level populations can be
seen by the inclusion
of higher-order driving field terms. A control of the level population can be obtained for the higher-order driving field terms.
Also the inclusion of higher-order terms in the driving field improves
the purity of the system. We can also see the effect of the Lindblad parameter $\Gamma$ on the purity of the system. Enhancement in purity is observed as dissipation in
the system is reduced. Also we have shown that the evolution of entropy of the system gives complementary information. We have also investigated carefully
the quantum state evolution of the NV spin in the presence of a spin-bath comprised of $^{13}$C nuclei. The CCD schemes does well in protecting the
state of the system in the presence of random fields originating from the spin-bath and the fluctuations due to the driving field itself.
In particular, the CCD driving protocol minimizes the entropy of the system and maximizes purity.

\section{Acknowledgment}

One of us (ARPR) thanks the Alexander von Humboldt Stiftung for support under a Wiedereinladung.
Financial support by the Deutsche Forschungsgemeinschaft (DFG) through SFB 762 is
gratefully acknowledged. SKM acknowledges Department of Science and Technology, India for support under the grant of INSPIRE Faculty Fellowship.
\appendix

\section{CCD scheme for environment and driving field fluctuations: A case study for two-level system}
\label{append1}
Let us consider a two-level system in contact with the environment. The Hamiltonian of the system is:
\begin{eqnarray}
 H_0=\frac{\omega+\zeta_{b}(t)}{2}\sigma_z
\end{eqnarray}
The role of the spin-bath environment is to modify the energy gap between the two eigenstates $\vert\uparrow\rangle$ and
 $\vert\downarrow\rangle$ of the Hamiltonian. In the above equation $\zeta_{b}(t)$ is a noise term due to the
spin-bath environment. In order to suppress the magnetic noise, we employ a continuous periodic driving field. However,
the driving field may also contain fluctuations. Let us consider a first-order driving field term given by    
\begin{eqnarray}
 H_d^{(1)}=\Omega_1(1+\zeta_1(t))\cos(\omega t)\sigma_x.
\end{eqnarray}
In the interaction picture with respect to $H_0^1=\frac{\omega}{2}\sigma_z$ and subsequently using
rotating wave approximation, the effective Hamiltonian takes the form
\begin{eqnarray}
 H_1^{(1)}=\frac{\Omega_1 (1+\zeta_1(t))}{2}\sigma_x +\frac{\zeta_b(t)}{2}\sigma_z.
\end{eqnarray}
The dressed states are the eigenstates of the Pauli matrix $\sigma_x$, i.e.,
 $\vert\rightarrow\rangle_x=\frac{1}{\sqrt{2}}(\vert\uparrow\rangle+\vert\downarrow\rangle)$ and $\vert\leftarrow\rangle_x=\frac{1}{\sqrt{2}}(\vert\uparrow\rangle-
\vert\downarrow\rangle)$. It can be argued here that the magnetic noise term affects the transition between $\vert\uparrow\rangle$ and $\vert\downarrow\rangle$
while the dressed states are protected against $\zeta_b(t)$. In this manner we could decouple the system from the environment.
However, the fluctuations $\zeta_1(t)$ in the driving field still modify the rate of transition between dressed states. In order to suppress $\zeta_1(t)$, we
apply a second order driving field term given by
\begin{eqnarray}
 H_d^{(2)}=\Omega_2(1+\zeta_2(t))\cos(\omega t+\frac{\pi}{2})\cos(\Omega_1t)\sigma_x
\end{eqnarray}
Using this second order term in the total Hamiltonian and considering interaction picture with respect to $H_0^1=\frac{\omega}{2}\sigma_z$ followed by the
rotating wave approximation, the effective Hamiltonian
turns out to be
\begin{eqnarray}
 H_1^{(1)}&=&\frac{\Omega_1 (1+\zeta_1(t))}{2}\sigma_x +\Omega_2(1+\zeta_2(t))\cos(\Omega_1t)\sigma_y \nonumber \\
&+&\frac{\zeta_b(t)}{2}\sigma_z.
\end{eqnarray}
 Let us take it further to second order interaction picture with respect to $H_0^{(2)}=\frac{\omega}{2}\sigma_x$. The effective Hamiltonian in the second order interaction picture will take the form
\begin{eqnarray}
  H_1^{(2)}=\frac{\Omega_2 (1+\zeta_2(t))}{2}\sigma_y+\frac{\Omega_1\zeta_1(t)}{2}.
\end{eqnarray}
The second order dressed states are eigenstates of the Pauli matrix $\sigma_y$, that is,
$\vert\rightarrow\rangle_y=\frac{1}{\sqrt{2}}(\vert\uparrow\rangle+i\vert\downarrow\rangle)$ and $\vert\leftarrow\rangle_y=\frac{1}{\sqrt{2}}(\vert\uparrow\rangle-
i\vert\downarrow\rangle)$. These second order dressed states are protected against fluctuations $\zeta_1(t)$ due the first order driving field. In
the procedure discussed here we have adopted rotating wave approximation which requires $\Omega_2\ll\Omega_1$.

\end{document}